\documentclass[aps,prd,superscriptaddress,showpacs,preprint]{revtex4}
\usepackage{graphicx, bm}
\begin{document}
\draft
\title{Bounding the electromagnetic and weak dipole moments of the tau-lepton
in a simplest little Higgs model}

\author{ A. Guti\'errez-Rodr\'{\i}guez}
\affiliation{\small Facultad de F\'{\i}sica, Universidad Aut\'onoma de Zacatecas\\
         Apartado Postal C-580, 98060 Zacatecas, M\'exico.\\ }

\date{\today}

\begin{abstract}

From the total cross section for the reaction $e^+e^-\to
\tau^+\tau^-\gamma$ at the $Z_1$ pole and in the framework of a
simplest little Higgs model (SLHM), we get a limit on the
characteristic energy scale of the model $f$, $f \geq 5.4$ $TeV$,
which in turn induces bounds on the electromagnetic and weak
dipole moments of the tau-lepton. Our bounds on the
electromagnetic moments are consistent with the bounds obtained by
the L3 and OPAL collaborations for the reaction $e^+e^-\to
\tau^+\tau^-\gamma$. We also obtained bounds on the tau weak
dipole moments which are consistent with the bounds obtained
recently by the DELPHI and ALEPH collaborations from the reaction
$e^+e^-\to \tau^+\tau^-$.
\end{abstract}


\vspace{5mm}

\maketitle

\section{Introduction}

The production of tau-lepton pairs in high energy  $e^+e^-$
collisions has been used to set bounds on its electromagnetic and
weak dipole moments \cite{Lohmann,L3D,DELPHI,ALEPH}. In the
Standard Model (SM) \cite{S.L.Glashow,S.Weinberg,A.Salam}, the
$\tau$ anomalous magnetic moment (MM) $a_\tau=(g_\tau-2)/2$ is
predicted to be $(a_\tau)_{SM}=0.0011773(3)$ \cite{Samuel,Hamzeh}
and the respective electric dipole moment (EDM) $d_\tau$ is
generated by the GIM mechanism only at very high order in the
coupling constant \cite{Barr}. Similarly, the weak MM and EDM are
induced in the SM at the loop level giving
$a^W_\tau=-(2.10+0.61i)\times 10^{-6}$ \cite{Bernabeu,Bernabeu1}
and $d^W_\tau \leq 8\times 10^{-34}e$cm \cite{Bernreuther,Booth}.
Since the current  bounds on these dipole moments
\cite{Lohmann,L3D,DELPHI,ALEPH} are well above the SM predictions,
it has been pointed out that these quantities are excellent
candidates to look for physics beyond the SM \cite{Bernabeu,
Bernabeu1,Bernreuther,Booth,Gonzalez-Garcia,Poulose,Huang,Escribano,
Grifols,Taylor,G.Gonzalez,G.Gonzalez1,G.Gonzalez2}. The couplings
of the photon and $Z$ gauge boson to charged leptons may be
parameterized in the following form:

\begin{equation}
\Gamma^{\alpha}_V=eF_{1}(q^{2})\gamma^{\alpha}+\frac{ie}{2m_l}F_{2}(q^{2})\sigma^{\alpha
\mu}q_{\mu}+ eF_{3}(q^{2})\gamma_5\sigma^{\alpha \mu}q_{\mu},
\end{equation}

\noindent where $V=\gamma, Z$, $m_l$ is the lepton mass and
$q=p'-p$ is the momentum transfer. The $q^2$-dependent
form-factors $F_i(q^2)$ have familiar interpretations for $q^2=0$:
$F_1(0)\equiv Q_l$ is the electric charge; $F_2(0)\equiv a_l$; and
$F_3\equiv d_l /Q_l$. The weak dipole moments are defined in a
similar way: $F^Z_2(q^2=m^2_Z)=a^W_\tau$ and
$F^Z_3(q^2=m^2_Z)=d^W_\tau/e$. The measurement of $a^W_\tau$ and
$d^W_\tau$ has been done in the $Z_1 \to \tau^+ \tau^-$ decay mode
at LEP. The latest bounds obtained for the electromagnetic and
weak dipole moments from the DELPHI and ALEPH collaborations at
the 95$\%$ C.L. are: $-0.052 < a_\tau < 0.013$, $-0.22 < d_\tau
(10^{-16}\hspace{1mm} e\mbox{cm}) < 0.45$ and $a^W_\tau <
1.1\times 10^{-3}$, $d^W_\tau < 0.50\times 10^{-17}e\mbox{cm}$
\cite{DELPHI,ALEPH}.

The first limits on the MM and EDM of the $\tau$ lepton were
obtained by Grifols and M\'endez using L3 data \cite{Grifols}:
$a_\tau\leq 0.11$ and $d_\tau\leq 6\times 10^{-16}e$cm. Escribano
and Mass\'o \cite{Escribano} later used electroweak precision
measurements to get $d_\tau \leq 1.1\times 10^{-17}e$cm and
$-0.004\leq a_\tau \leq 0.006$ at the $2\sigma$ confidence level.
There is extensive theoretical work done in models beyond the SM
that contribute to EDM of charged leptons. In Ref. \cite{Iltan},
the EDM of charged leptons are studied assuming that they have
Gaussian profiles in extra dimensions. In \cite{Dutta} the lepton
EDM has been analyzed in the framework of the seesaw model. The
electric dipole moments of the leptons in the version III of the
2HDM are considered in \cite{Iltan1}. The work \cite{Iltan2} was
related to the lepton EDM in the framework of the SM with the
inclusion of non-commutative geometry. Furthermore, the effects of
non-universal extra dimensions on the EDM of fermions in the two
Higgs doublet model have been estimated in Ref. \cite{Iltan3}. In
\cite{A.Gutierrez1,A.Gutierrez2}, limits on the electromagnetic
and weak dipole moments of the tau-lepton in the framework of a
left-right symmetric model (LRSM) and a class of $E_6$ inspired
models with an additional neutral vector boson $Z_\theta$ have
been analyzed.

Theoretically, numerous new physics models are proposed with
different roles of Higgs in the models. The little Higgs model
(LHM) \cite{Arkani1,Arkani2} has been proposed for solving the
little hierarchy problem. In this scenario, the Higgs boson is
regarded as a pseudo Nambu-Goldstone boson associated with a
global symmetry at some higher scale. Though the symmetry is not
exact, its breaking is specially arranged to cancel quadratically
divergent corrections to the Higgs mass term at 1-loop level. This
is called the little Higgs mechanism. As a result, the scale of
new physics can be as high as 10 $TeV$ without a fine-tuning on
the Higgs mass term. In these models, relatively light Higgs boson
mass is due to its identity as a pseudo Goldstone boson of some
enlarged global symmetries. Among various little Higgs models, the
simplest little Higgs model (SLHM) \cite{Kaplan,Schmaltz,Dias} is
attractive due to its relatively simple theory structure. Detailed
discussions on SLHM can be found in the literature
\cite{Kaplan,Schmaltz,Dias}.

On the $Z_1$ peak, where a large number of $Z_1$ events are
collected at $e^{+}e^{-}$ colliders, one may hope to constrain or
eventually measure the electromagnetic and weak dipole moments of
the $\tau$ by selecting $\tau^{+}\tau^{-}$ events accompanied by a
hard photon. The Feynman diagrams which give the most important
contribution to the cross section from $e^{+}e^{-}\rightarrow
\tau^+ \tau^- \gamma$ are shown in Fig. 1. The total cross section
of $e^{+}e^{-}\rightarrow \tau^+ \tau^- \gamma$ will be evaluated
at the $Z_1$-pole in the framework of a simplest little Higgs
model. The numerical computation for the anomalous magnetic and
the electric dipole moments of the tau is done using the data
collected by the L3 and OPAL collaborations at LEP \cite{L3,OPAL}.
We are interested in studying the effects induced by the effective
couplings associated to the weak and electromagnetic moments of
the tau lepton given in Eq. (1). For this purpose, we will take
the respective anomalous vertices $\tau\tau\gamma$ and $\tau\tau
Z_1$, one at the time, in diagrams (1) and (2) of Fig. 1. The
numerical computation for the respective transition amplitudes
will be done using the data collected by these collaborations.

Our aim in this paper is to analyze the reaction
$e^{+}e^{-}\rightarrow \tau^+ \tau^- \gamma$ in the $Z_1$ boson
resonance. The analysis is carried out in the context of a
simplest little Higgs model \cite{Kaplan,Schmaltz,Dias} and we
attribute electromagnetic and weak dipole moments to the tau
lepton. Processes measured in the resonance serve to set limits on
the tau electromagnetic and weak dipole moments. First, using as
an input the results obtained by the L3 and OPAL collaborations
\cite{L3,OPAL} for the tau MM and EDM in the process
$e^{+}e^{-}\rightarrow \tau^+ \tau^- \gamma$, we will set a limit
on the SLHM energy scale $f$ which is similar to that obtained
through oblique corrections \cite{Marandella}, as well as that
obtained recently from the $Z_1$ leptonic decay \cite{Dias}. We
then use this limit on $f$ to get bounds on the weak dipole
moments of the tau from the same L3/OPAL data. We have found that
these limits are consistent with the new bounds obtained by the
DELPHI and ALEPH collaborations from the process $e^+e^-\to
\tau^+\tau^-$ \cite{L3E,DELPHI,ALEPH}.

This paper is organized as follows: In Sect. II we present the
calculation of the cross section for the process
$e^{+}e^{-}\rightarrow \tau^+ \tau^- \gamma$ in a simplest little
Higgs model. In Sect. III we present our results for the numerical
computations and, finally, we present our conclusions in Sect. IV.

\section{The Total Cross Section}

In this section we calculate the total cross section for the
reaction $e^{+}e^{-}\rightarrow \tau^+\tau^- \gamma$ using the
neutral current lagrangian given in Eq. (22) of Ref. \cite{Dias}
for the SLHM for diagrams 1 and 2 of Fig. 1. The respective
transition amplitudes are thus given by

\begin{eqnarray}
{\cal M}_{1}&=&\frac{-g^{2}}{4\cos^{2}\theta_{W}(l^{2}-m^{2}_{\tau})}[\bar u(p_{3})\Gamma^{\alpha}(l\llap{/}+m_{\tau})\gamma^{\beta}(g^{\tau}_V-g^{\tau}_A \gamma_{5})v(p_{4})]\nonumber\\
         &&\frac{(g_{\alpha\beta}-p_{\alpha}p_{\beta}/M^{2}_{Z_1})}{[(p_{1}+p_{2})^{2}-M^{2}_{Z_1}-i\Gamma^{2}_{Z_1}]}[\bar u(p_{2})\gamma^{\alpha}(g^{e}_V-g_A^{e}\gamma_{5})v(p_{1})]\epsilon^{\lambda}_{\alpha},
\end{eqnarray}

\begin{eqnarray}
{\cal M}_{2}&=&\frac{-g^{2}}{4\cos^{2}\theta_{W}(k^{2}-m^{2}_{\tau})}[\bar u(p_{3})\gamma^{\beta}(g^{\tau}_V-g^{\tau}_A\gamma_{5})(k\llap{/}+m_{\tau})\Gamma^{\alpha}   v(p_{4})]\nonumber\\
         &&\frac{(g_{\alpha\beta}-p_{\alpha}p_{\beta}/M^{2}_{Z_1})}{[(p_{1}+p_{2})^{2}-M^{2}_{Z_1}-i\Gamma^{2}_{Z_1}]}[\bar u(p_{2})\gamma^{\alpha}(g_V^{e}-g_A^{e}\gamma_{5})v(p_{1})]\epsilon^{\lambda}_{\alpha},
\end{eqnarray}

\noindent where $\Gamma^{\alpha}$ is the tau-lepton
electromagnetic vertex which is defined in Eq. (1), while
$\epsilon^{\lambda}_{\alpha}$ is the polarization vector of the
photon. $l$ ($k$) stands for the momentum of the virtual tau
(antitau), and the coupling constants $g^{l}_V$ and $g^l_A$ with
$l=e, \mu, \tau$ are given in Eq. (23) of Ref. \cite{Dias}.

The MM, EDM and the characteristic energy scale of the simplest
little Higgs model $f$ give a contribution to the differential
cross section for the process $e^{+}e^{-}\rightarrow
\tau^+\tau^-\gamma$ of the form:

\begin{eqnarray}
\sigma(e^{+}e^{-}\rightarrow \tau^+\tau^-\gamma)&=&\int
\frac{\alpha^2}{48\pi} [\frac{e^2a^2_\tau}{4m^2_\tau}+d^2_\tau]
[\frac{1-4x_W+8x^2_W}{x^2_W(1-x_W)^2}]\nonumber\\
&&[\frac{(1-4x_W+x^2_W)
(s-2\sqrt{s}E_\gamma)+\frac{1}{2}E^2_\gamma\sin^2\theta_\gamma}{(s-M^2_{Z_{1}})^2+M^2_{Z_{1}}\Gamma^2_{Z_{1}}}]\nonumber\\
&&[1+\frac{1}{8}\frac{(3-4x_W)}{(1-x_W)^2}(\frac{v^2}{f^2})]^4{E_\gamma
dE_\gamma d\cos\theta_\gamma},
\end{eqnarray}

\noindent where $x_{W}\equiv \sin^{2}\theta_{W}$, $v$ is the
vacuum expectation value and $E_{\gamma}$, $\cos\theta_{\gamma}$
are the energy and the opening angle of the emmited photon.

It is useful to consider the smallness of the factor
$(\frac{v^2}{f^2})$, to approximate the cross section in Eq. (4)
by its expansion in powers of $(\frac{v^2}{f^2})$ to the linear
term:
$\sigma=(\frac{e^2a^2_\tau}{4m^2_\tau}+d^2_\tau)[A+B(\frac{v^2}{f^2})+O((\frac{v^2}{f^2})^2
)]$, where $A$ and $B$ are constants which can be evaluated. Such
an approximation for deriving the bounds of $a_\tau$ and $d_\tau$
is more illustrative and easier to manipulate.

For $(\frac{v^2}{f^2})< 1$, the total cross section for the
process $e^{+}e^{-}\rightarrow \tau^+\tau^-\gamma$ is given by

\begin{equation}
\sigma(e^{+}e^{-}\rightarrow
\tau^+\tau^-\gamma)=(\frac{e^2a^2_\tau}{4m^2_\tau}+d^2_\tau)[A+B(\frac{v^2}{f^2})+O((\frac{v^2}{f^2})^2
)],
\end{equation}

\noindent where $A$ explicitly is

\begin{eqnarray}
A&=&\int \frac{\alpha^2}{48\pi}
[\frac{1-4x_W+8x^2_W}{x^2_W(1-x_W)^2}]\nonumber\\
&&[\frac{(1-4x_W+x^2_W)
(s-2\sqrt{s}E_\gamma)+\frac{1}{2}E^2_\gamma\sin^2\theta_\gamma}{(s-M^2_{Z_{1}})^2+M^2_{Z_{1}}\Gamma^2_{Z_{1}}}]
{E_\gamma dE_\gamma d\cos\theta_\gamma},\nonumber\\
\end{eqnarray}

\noindent while $B$ is given by

\begin{eqnarray}
B&=&\int \frac{\alpha^2}{96\pi}
[\frac{1-4x_W+8x^2_W}{x^2_W(1-x_W)^2}]\nonumber\\
&&[\frac{(1-4x_W+x^2_W)
(s-2\sqrt{s}E_\gamma)+\frac{1}{2}E^2_\gamma\sin^2\theta_\gamma}{(s-M^2_{Z_{1}})^2+M^2_{Z_{1}}\Gamma^2_{Z_{1}}}]\nonumber\\
&&[\frac{(3-4x_W)}{(1-x_W)^2}(\frac{v^2}{f^2})]{E_\gamma dE_\gamma
d\cos\theta_\gamma}.
\end{eqnarray}

\noindent The expression given for $A$ corresponds to the cross
section previously reported by Grifols and M\'endez for the SM
\cite{Grifols}, while $B$ comes from the contribution of the SLHM.
Evaluating the limit when the characteristic energy scale $f \to
\infty$, the second term in (5) is zero and Eq. (5) is reduced to
the expression (4) given in Ref. \cite{Grifols}.

In the case of the weak dipole moments, to get the expression for
the differential cross section, we have to substitute  the $Z_1$
SLHM couplings given in Eq. (23) of Ref. \cite{Dias} with the
respective weak dipole moments included in Eq. (1), that is to say
$a^W_\tau=F^Z_2(q^2=m^2_Z)$ and $d^W_\tau=eF^Z_3(q^2=m^2_Z)$. We
do not reproduce the analytical expressions here because they are
rather similar to the term given in Eq. (4). In the following
section we will present the bounds obtained for the tau dipole
moments using the data published by the L3 and OPAL collaborations
for the reaction $e^{+}e^{-}\rightarrow \tau^+\tau^-\gamma$
\cite{L3,OPAL}.

\section{Results}

In practice, detector geometry imposes a cut on the photon polar
angle with respect to the electron direction, and further cuts
must be applied on the photon energy and minimum opening angle
between the photon and tau in order to suppress background from
tau decay products. In order to evaluate the integral of the total
cross section as a function of the parameters of the SLHM, that is
to say, $f$, we require cuts on the photon angle and energy to
avoid divergences when the integral is evaluated at the important
intervals of each experiment. We integrate over
$\cos\theta_\gamma$ from $-0.74$ to $0.74$ and $E_\gamma$ from 5
$GeV$ to 45.5 $GeV$ for various fixed values of the characteristic
energy scale $f=(1.7, 5.2, 5.4, 5.6, 7, 10)$ $TeV$  (as
illustrated in Fig. 2) according to Refs. \cite{Marandella,Dias}.
Using the numerical values $\sin^2\theta_W=0.2314$,
$M_{Z_1}=91.18$ $GeV$, $\Gamma_{Z_1}=2.49$ $GeV$ and $m_\tau=
1.776$ $GeV$, we obtain the cross section $\sigma=\sigma(f,
a_\tau,d_\tau)$.

In Fig. 2, we show the dependence of the total cross section for
the process $e^+e^-\to \tau^+\tau^-\gamma$ with respect to the
SLHM energy scale $f$. Using the data $\sigma=(1.472\pm 0.006 \pm
0.020)$ $nb$ Refs. \cite{Taylor,L3} for the cross section, where
the first error is statistical and the second is systematic, we
get the following limit for $f$:

\begin{equation}
f \geq 5.4\hspace{2mm} TeV,
\end{equation}

\noindent which is consistent with that obtained through oblique
corrections \cite{Marandella}, as well as that obtained recently
from an analysis on the decay width $\Gamma(Z_1 \to e^+e^-)$
\cite{Dias}.

As was discussed in Ref. \cite{L3}, $N\approx\sigma(f, a_\tau,
d_\tau){\cal L}$, using Poisson statistic \cite{L3,Barnett}, we
require that $N\approx\sigma(f, a_\tau, d_\tau){\cal L}$ be less
than 1559, with ${\cal L}= 100$ $pb^{-1}$, according to the data
reported by the L3 collaboration Ref. \cite{L3} and references
therein. Taking this into consideration, we can get a bound for
the tau magnetic moment as a function of $f$ with $d_\tau=0$. The
values obtained for this bound for several values of $f$ are
included in Table 1. The previous analysis and comments can
readily be translated to the EDM of the tau with $a_\tau=0$. The
resulting bounds for the EDM as a function of $f$ are shown in
Table 1. As expected, the limits obtained for the electromagnetic
dipole moments of the tau lepton are consistent with those
obtained by these collaborations from the data for the process
$e^+e^-\to \tau^+\tau^-\gamma$ \cite{L3,OPAL}.

\begin{center}
\begin{tabular}{|c|c|c|}\hline
$f(TeV)$&${a_\tau}$&$d_\tau(10^{-16}e \mbox{cm})$\\
\hline \hline
$1.7$  & 0.0490 & 2.800\\
\hline
$5.2$  & 0.0511 & 2.855\\
\hline
$5.4$  & 0.0512 & 2.856\\
\hline
$5.6$  & 0.0513 & 2.857\\
\hline
$7$    & 0.0514 & 2.858\\
\hline
$10$   & 0.0515 & 2.860\\
\hline
\end{tabular}
\end{center}

\begin{center}
Table 1. Bounds on the $a_\tau$ MM and $d_\tau$ EDM of the
$\tau$-lepton for different values of the characteristic energy
scale of the model $f$. We have applied the cuts used by L3 for
the photon angle and energy.
\end{center}

\vspace{3mm}

The bounds for the weak dipole moments of the tau-lepton according
to the data from the L3 and OPAL collaborations \cite{L3,OPAL} for
the energy and the opening angle of the photon, as well as the
luminosity and the event numbers, are given in the Table 2. As we
can see, the use of the limit obtained for the $f$ characteristic
energy scale of the model also induces bounds for the tau weak
dipole moments, which are already consistent with those bounds
recently obtained by the DELPHI and ALEPH collaborations in the
process $ e^+e^-\to \tau^+\tau^-$ \cite{DELPHI,ALEPH}. Our results
in Table 2 for $f=(1.7, 5.2, 5.4, 5.6, 7, 10)$ $TeV$ \cite{Dias}
differ by about a factor of two of the bounds obtained by the
DELPHI and ALEPH collaborations for the weak tau dipole moments
\cite{Lohmann,DELPHI,ALEPH}; our analysis is not sensitive to the
real and imaginary parts of these parameters separately. In order
to improve these limits it might be necessary to study direct
CP-violating effects \cite{M.A.Perez,Larios}.

\begin{center}
\begin{tabular}{|c|c|c|}\hline
$f(TeV)$ & ${a^W_\tau(10^{-3}) }$ &$d^W_\tau(10^{-17}e \mbox{cm})$\\
\hline \hline
$1.7$ & 2.020 & 1.121\\
\hline
$5.2$ & 2.052 & 1.138\\
\hline
$5.4$ & 2.053 & 1.139\\
\hline
$5.6$ & 2.054 & 1.140\\
\hline
$7$ & 2.055 & 1.141\\
\hline
$10$ & 2.056 & 1.142\\
\hline
\end{tabular}
\end{center}

\begin{center}
Table 2. Bounds on the $a^W_\tau$ anomalous weak MM and $d^W_\tau$
weak EDM of the $\tau$-lepton for different values of the
characteristic energy scale of the model $f$. We have applied the
cuts used by L3 for the photon angle and energy.
\end{center}

We plot the total cross section in Fig. 3 as a function of the
characteristic energy scale of the model $f$ for the bounds of the
magnetic moment given in Table 1. In Fig. 3, for $f=10$ $TeV$ we
reproduce the data previously reported in the literature. Our
results for the dependence of the differential cross section on
the photon energy versus the cosine of the opening angle between
the photon and the beam direction ($\theta_\gamma$) are presented
in Fig. 4 for $f = 5.4$ $TeV$ and $a_\tau = 0.0512$. Besides we
plot the differential cross section in Fig. 5 as a function of the
photon energy for the bounds of the magnetic moments given in
Table 1. We observe in this figure that the energy distributions
are consistent with those reported in the literature. Finally, we
find that the effects induced by the tree level $Z_1e^+e^-$ and
$Z_1\tau^+\tau^-$ couplings in the SLHM increase the cross section
of the process $e^+e^- \to \tau^+\tau^-\gamma$, and the
predictions on the electromagnetic and weak dipole moments of the
tau lepton are better estimated. However, it is necessary to make
an analysis at loop level for the process $e^+e^- \to
\tau^+\tau^-\gamma$ in the context of a little Higgs model.

\section{Conclusions}

We have determined limits on the electromagnetic and weak dipole
moments of the tau-lepton using the data published by the L3 and
OPAL collaborations for the $e^+e^-\to \tau^+\tau^-\gamma$ at the
$Z_1$ pole. We were able to get limits on the weak dipole moments
by constraining the SLHM energy scale $f$ from the electromagnetic
dipole moments obtained by these collaborations. We then used this
limit to determined bounds on the electromagnetic and weak dipole
moments of the tau lepton using the data published by the L3 and
OPAL collaborations for the process $e^+e^-\to
\tau^+\tau^-\gamma$. In addition, we obtained bounds for the weak
dipole moments similar to those obtained recently by the
DELPHI/ALEPH collaborations from the process $e^+e^-\to
\tau^+\tau^-$ \cite{DELPHI,ALEPH}. In particular, from the limit
$f \to \infty$, our bound take the value previously reported in
Ref. \cite{Grifols} for the SM. As far as the weak dipole moments
are concerned, our limits given in Tables 1 and 2 are consistent
with the experimental bounds obtained at LEP with the two-body
decay mode $Z_1 \to \tau^+ \tau^-$ \cite{Lohmann}. In addition,
the analytical and numerical results for the total cross section
have never been reported in the literature before and could be of
relevance for the scientific community.

\vspace{1cm}

\begin{center}
{\bf Acknowledgments}
\end{center}

We acknowledge support from CONACyT, SNI and PROMEP(M\'exico).

\newpage

\begin{figure}[t]
\centerline{\scalebox{0.8}{\includegraphics{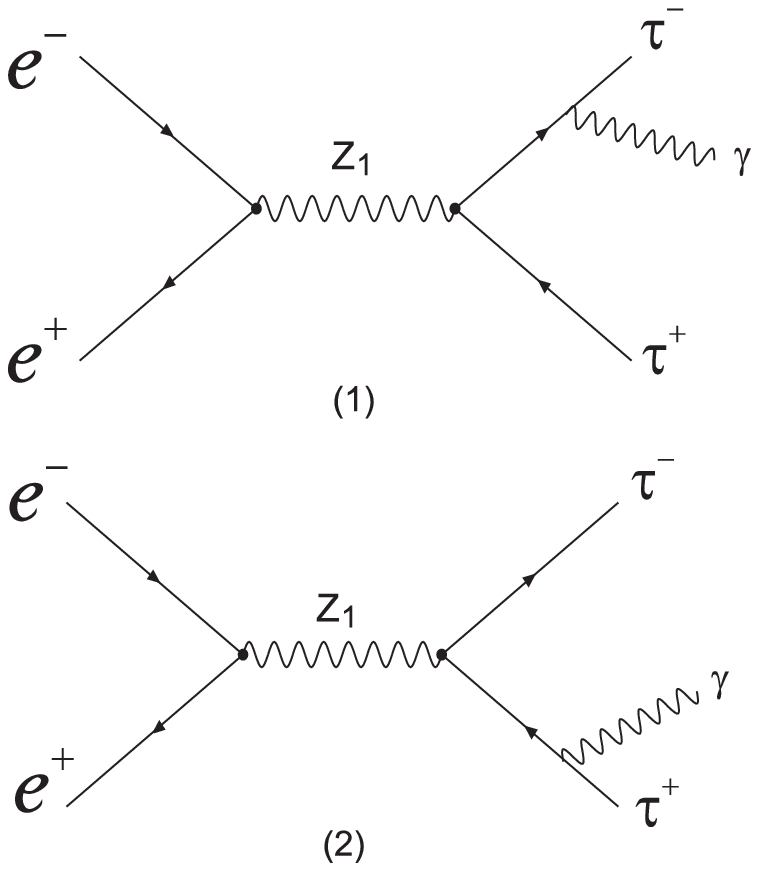}}}
\caption{ \label{fig:gamma} The Feynman diagrams contributing to
the process $e^{+}e^{-}\rightarrow \tau^+\tau^-\gamma$ in a
simplest little Higgs model.}
\end{figure}

\begin{figure}[t]
\centerline{\scalebox{0.75}{\includegraphics{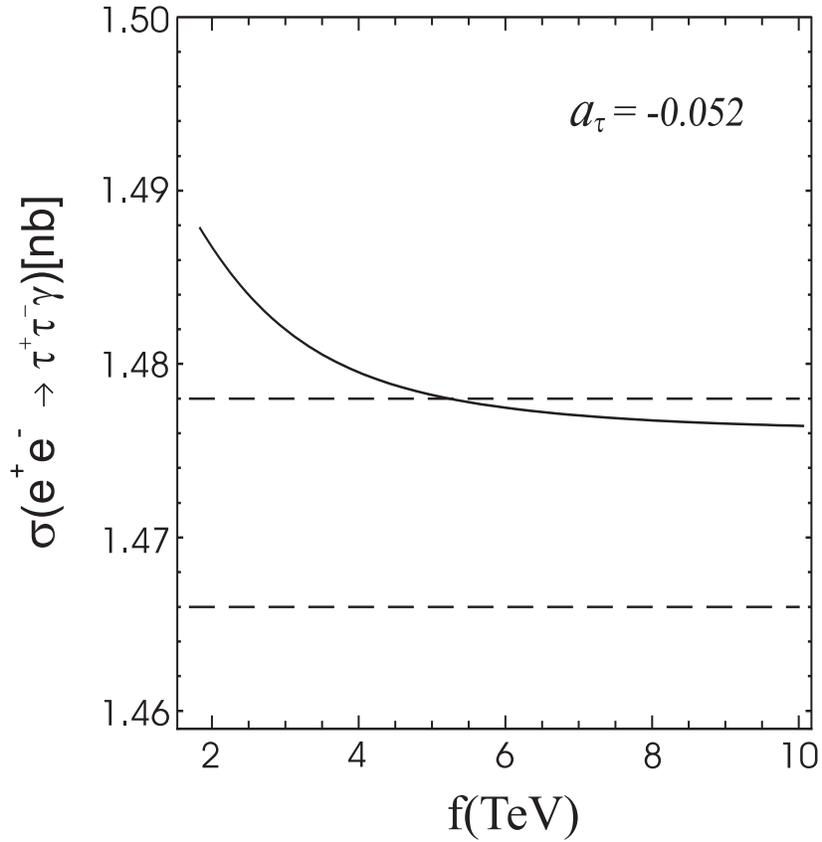}}}
\caption{ \label{fig:gamma} The curves show the shape for
$\sigma(e^{+}e^{-}\rightarrow \tau^+\tau^-\gamma)$ as a function
of the energy scale $f$.}
\end{figure}

\begin{figure}[t]
\centerline{\scalebox{0.9}{\includegraphics{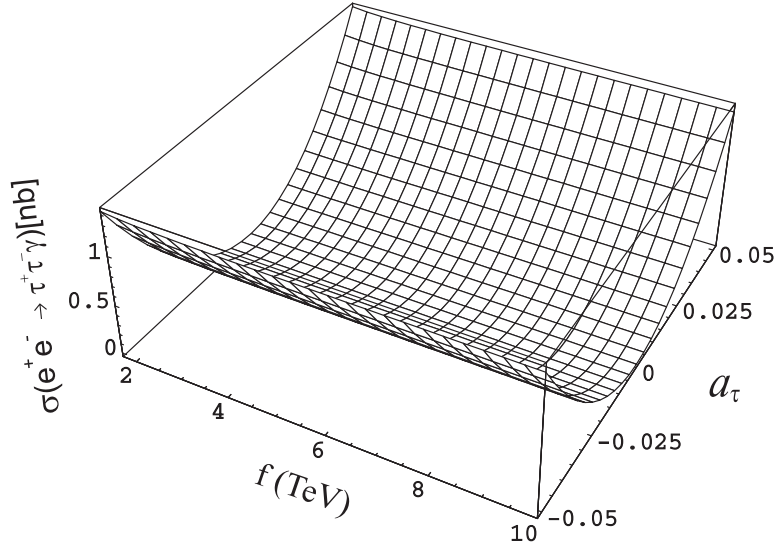}}}
\caption{ \label{fig:gamma} The total cross-section for
$e^{+}e^{-}\rightarrow \tau^+\tau^-\gamma$ as a function of the
characteristic energy scale $f$ and $a_\tau$ (Table 1).}
\end{figure}

\begin{figure}[t]
\centerline{\scalebox{0.75}{\includegraphics{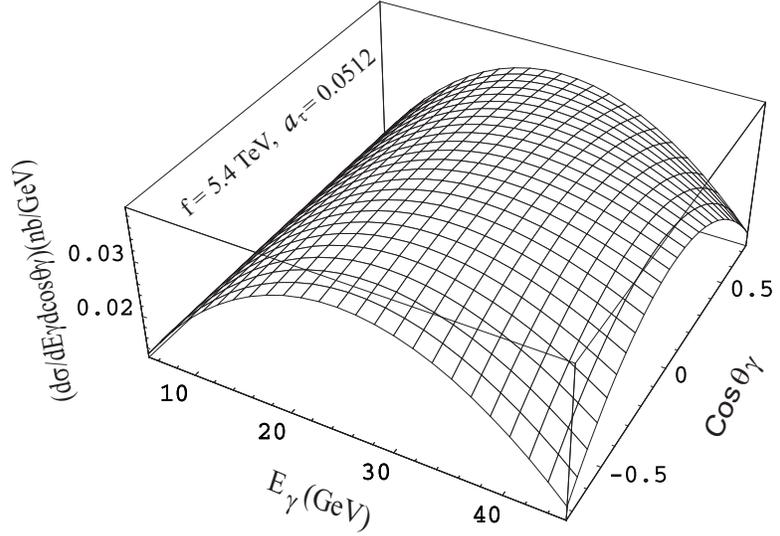}}}
\caption{ \label{fig:gamma} The differential cross section for
$e^{+}e^{-}\rightarrow \tau^+\tau^-\gamma$ as a function of
$E_\gamma$ and $\cos\theta_\gamma$ for $f=5.4$ $TeV$ and
$a_\tau=0.0512$.}
\end{figure}

\begin{figure}[t]
\centerline{\scalebox{0.8}{\includegraphics{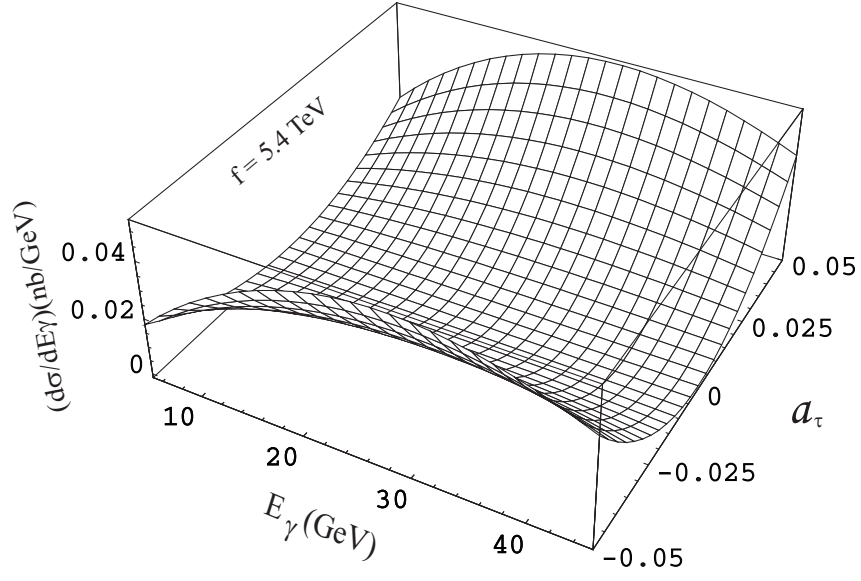}}}
\caption{ \label{fig:gamma} The differential cross section for
$e^{+}e^{-}\rightarrow \tau^+\tau^-\gamma$ as a function of
$E_\gamma$ and $a_\tau$ with $f=5.4$ $TeV$.}
\end{figure}

\end{document}